\documentclass[twocolumn,amssymb,showpacs,amsmath,nobibnotes,nofootinbib,aps,prb,superscriptaddress]{revtex4}

\usepackage{graphicx}
\usepackage{dcolumn}
\usepackage{bm}
\begin{document}
\preprint{APS/123-QED}
\title{Magnetic phases in the Kagom\'{e} staircase compound $\rm Co_3V_2O_8$}
\newcommand{\NVO}{$\rm Ni_3V_2O_8$}
\newcommand{\CVO}{$\rm Co_3V_2O_8$}
\newcommand{\MVO}{$\rm M_3V_2O_8$}
\newcommand{\afm}{anriferromagnetic}
\newcommand{\Kag}{Kagom\'{e}}
\author{N. R. Wilson}
\email{Nicola.R.Wilson@warwick.ac.uk}
\author{O. A. Petrenko}
\affiliation{Department of Physics, University of Warwick, Coventry, CV4 7AL, UK}
\author{L. C. Chapon}
\affiliation{ISIS Facility, Rutherford Appleton Laboratory, Chilton, Didcot, OX11 0QX, UK}
\date{\today}

\begin{abstract}
The low temperature properties of the \Kag -type system \CVO\ have been studied by powder neutron diffraction both in zero field and in applied magnetic field of up to 8~T. Below 6~K, the zero-field ground state is ferromagnetic with the magnetic moments aligned along the \textit{a}-axis. The size of the moment on one of the two Co sites, the so called cross-tie site, is considerably reduced compared to the fully polarized state. The application of a magnetic field in this phase is found to rapidly enhance the cross-tie site magnetic moment, which reaches the expected value of $\sim$~3~$\mu_{\rm B}$ by the maximum applied field of 8~T. Different reorientation behaviors are found for the Co cross-tie and spine sites, suggesting a more pronounced easy-axis anisotropy for moments on the spine sites. Rietveld refinements reveal that a simple model, where the spins on both cross-tie and spine sites rotate in the \textit{ac}-plane in a magnetic field, reproduces the experimental diffraction patterns well. In addition, it is found that at higher temperatures and moderate magnetic fields, the incommensurate antiferromagnetic order, corresponding to a transverse sinusoidal modulation above 8~K, is suppressed to be replaced by ferromagnetic order.
\end{abstract}

\maketitle

\section{Introduction}
A Heisenberg antiferromagnet on the two dimensional \Kag\ lattice (a net of corner sharing triangles) with only nearest neighbor interactions has an infinitely large number of classical ground states. The system remains a disordered spin liquid down to zero temperature, reflecting this macroscopic degeneracy. Recently, a new class of frustrated magnets, \Kag\ staircase oxides, have become a central point of intensive investigations.\cite{Chen, Szymczak, Kenzelmann, Rogado1, Lawes, Lawesferro} These compounds, of formula \MVO\ (M=Ni, Co, Cu or Zn) present an exchange topology closely related to the \Kag\ lattice. In these staircase oxides, layers of edge-sharing $\rm MO_6$ octahedra are separated by non-magnetic $\rm VO_4$ tetrahedra. The specific chemical bonding in these materials creates layers of magnetic Co$^{2+}$ ions (the V$^{5+}$ ions are nonmagnetic) buckled into a staircase formation -- hence the so-called \Kag\ staircase lattice.\cite{Rogado1} The magnetic layers have lower symmetry than that found in the exact \Kag\ lattices which, combined with next-nearest neighbor interactions and possible anisotropic exchange, partially releases the geometrical frustration.

These systems have attracted considerable interest because of the complex sequences of magnetic transitions observed. In particular, the magnetically ordered states for the Ni and Co analogs indicate strongly competing interactions. On lowering the temperature, the Ni system presents two incommensurate phases followed by a commensurate phase with remnant high temperature structures, and a final low-temperature commensurate phase.\cite{Lawes} It was found that inversion symmetry is broken in the low temperature incommensurate cycloidal phase, giving rise to a macroscopic polar vector in a narrow temperature range.\cite{Lawesferro} This multiferroic behavior has been related to similar effects recently observed in several other frustrated systems such as TbMnO$_3$\cite{Harrisreview} and CuFeO$_2,$~\cite{KimuraCuFeO2} where ferroelectric order is induced by a symmetry breaking process associated with a non-collinear magnetic arrangement.

Very recently, Chen {\it et al.}~\cite{Chen} have reported a detailed investigation of the magnetic ground state of \CVO\ by powder and single crystal neutron diffraction. The investigation reveals that on cooling a progressive transition from an incommensurate magnetic order to a ferromagnetic state exists along with a number of lock-in transitions related to commensurate antiferromagnetic order at intermediate temperatures. Contrary to what has been reported based on bulk properties measurements,\cite{Szymczak} the magnetic behavior is totally different to what has been observed for the Ni counterpart, suggesting different strengths of the long-range interactions and the magnetic anisotropy.

We report here the results of a powder neutron diffraction study of the \afm\ incommensurate and ferromagnetic structures of \CVO\ as a function of temperature and magnetic field. The analysis of the data brings to light new aspects of the magnetic properties of this material. In particular, the large anisotropy and unsaturated moments in the ferromagnetic ground state. Our results for the commensurate magnetic phases in zero-field are in perfect agreement with those reported by Chen {\it et al.}~\cite{Chen} We further report the parameters extracted from Rietveld refinement of data in the high-temperature incommensurate region, showing a smooth variation of phase factors in the vicinity of the lock-in transition. In addition, neutron diffraction reveals that both incommensurate and ferromagnetic states are extremely sensitive to the application of a magnetic field. At 9~K, moderate magnetic fields destabilize the incommensurate magnetic state inducing a direct transition to a ferromagnetic state analog to that found in zero-field at low temperatures. A magnetic field applied in the low-temperature ferromagnetic state does not change the periodicity of the magnetic structure. However, the magnetic moment of the Co ion on the cross-tie site that is considerably reduced in the zero-field structure, shows a net increase, eventually saturating at the same value as the Co spine moment. High-field data at low temperatures can be explained by a spin-reorientation process predominantly in the \textit{ac}-plane, with different behaviors for the two Co sites.

\section{Experimental}
Polycrystalline samples were prepared from the starting materials CoO and {$\rm V_2O_5$}.\cite{Balakrishnan} The reagents were thoroughly mixed by grinding and then heated in air to $800^\circ$C for 24 hours. The compound was then ground and heated in air at $1050^\circ$C for 24~hours, followed by a final grinding. Around 6~g of powder were used for the neutron experiment. Medium resolution neutron powder diffraction experiments were conducted on the GEM (General Materials) diffractometer at the ISIS facility of the Rutherford Appleton Laboratory, UK. Diffraction patterns were collected in zero-field at half degree intervals over the phase transition temperatures, and then at degree intervals over the remaining temperatures between 2 and 12~K. The counting time for each run was approximately 1~hour. The diffraction patterns in magnetic fields were collected using a 10~T Oxford Instruments superconducting magnet. Data were recorded at constant temperatures of 2~K and 9~K with an increasing magnetic field of 0.5, 1.5, 2.5, 4.0 and 8.0~T. The counting time for each run was approximately 2~hours. For in-field experiments, the sample was tightly packed into a vanadium can to avoid reorientation of the crystallites. Rietveld refinements were performed with the FullProF program.~\cite{fullprof} Possible symmetry arrangements of the low-temperature magnetic structure were determined using representation analysis. The results, are presented using Kovalev's notation.\cite{Kovalev}

\section{Results}
\subsection{Zero-field magnetic structure}
\begin{figure}[tb]
\includegraphics[width=\columnwidth]{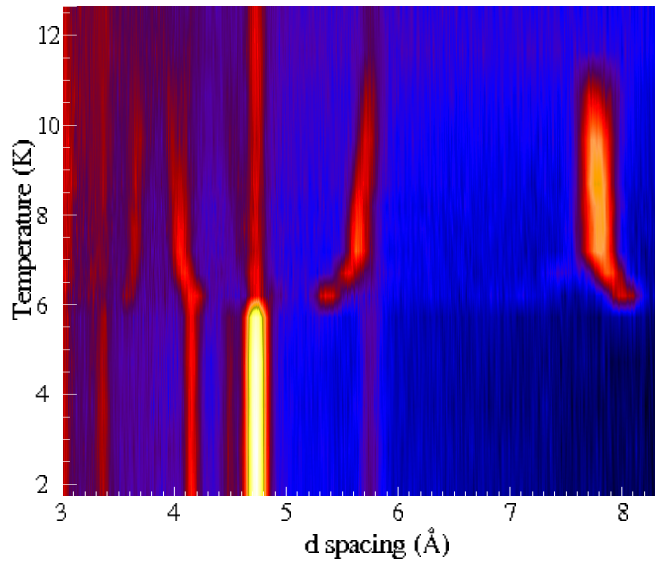}
\caption{\label{fig:thermodiffractogram} (color online) Zero field neutron diffraction pattern of \CVO\ as a function of temperature. Data from a bank of detectors situated at $2\theta=18^\circ$ are shown and the intensity is color coded.}
\end{figure}

\CVO\ was found to order around 11~K, in agreement with previous reports.\cite{Rogado1, Chen, Balakrishnan, Szymczak} On cooling, it shows a complex sequence of phase transitions as indicated by the temperature dependence of the neutron diffraction pattern, shown in Fig.~\ref{fig:thermodiffractogram}, clearly demonstrating abrupt shifts in the positions and relative intensities of magnetic Bragg peaks. First, incommensurate magnetic ordering sets in with a propagation vector $\textbf{k}=(0,\delta,0)$, where $\delta=0.54(1)$ at 9.7~K and decreases rapidly to lock at $\delta=\frac{1}{2}$ around 8.0~K. With further cooling, the structure becomes incommensurate again with a rapid decrease of $\delta$. At 5.7~K, a sudden lock-in transition to $\textbf{k}=(0,0,0)$ is evidenced, again in excellent agreement with recent results. 

The temperatures measured in each phase are as follows: The high temperature incommensurate phase (HT$_{\rm inc}$) includes 10.7, 9.7 and 8.7~K, the high temperature $\delta =\frac{1}{2}$ phase (HT$_{\delta =\frac{1}{2}}$) includes 8.2, 7.7 and 7.2~K, the intermediate phase where $\delta$ rapidly decreases includes 6.7 and 6.2~K, and the low temperature ferromagnetic phase (LT$_{\rm F}$) includes 5.7~K and below. Due to relatively large temperature steps (1~K) in our experiments, the exact temperature dependence including further lock-in transitions in the intermediate phase as reported by Chen {\it et al.}~\cite{Chen} could not be precisely mapped out. 

Rietveld analysis of the magnetic structure has been performed in the HT$_{\rm inc}$ phase, the HT$_{\delta =\frac{1}{2}}$ phase and in the LT$_{\rm F}$ $\textbf{k}=0$ phase. In the intermediate temperature range, however, the presence of magnetic Bragg peaks indexed with third order harmonics of the propagation vector complicates the analysis. In particular, the fundamental reflection is only partially observed in the d-spacing range accessible from our lowest scattering angle detectors; the large peak occurs at 22.8~\AA\ at 7.2~K and moves up to 25~\AA\ at 6.2~K.

\begin{figure}[tb]
\includegraphics[angle=-90,width=1\columnwidth]{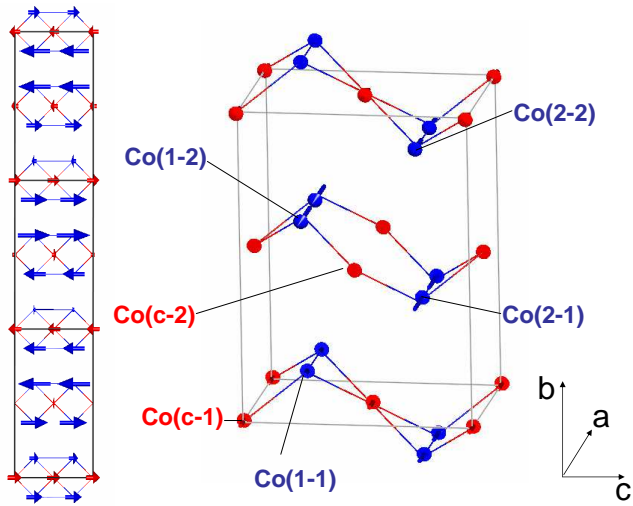}
\caption{\label{fig:structure} (color online) Incommensurate magnetic structure at 9.7~K. Only Co ions on the cross-tie sites and spine sites are shown in red (light grey) and blue (dark grey) respectively. Connections between first neighbor Co atoms are shown as thin solid lines. The right panel displays a single crystallographic unit-cell and labels the different atoms related by symmetry (See text for details). }
\end{figure}

\begin{table*}[tb]
\begin{tabular}{c|cccccc}
 & & & Atoms & & & \\
Irreps. & Co(c-1) & Co(c-2) & Co(1-1) & Co(1-2) & Co(2-1) & Co(2-2) \\
\hline $\Gamma_1$ & (u$_1$,0,0) & (-u$_1$,0,0)$\phi$ & (0,v$_2$,0) & (0,-v$_2$,0)$\phi$ & (0,v$_3$,0) & (0,-v$_3$,0)$\phi$ \\
$\Gamma_2$ & (0,v$_1$,w$_1$) & (0,v$_1$,-w$_1$)$\phi$ & (0,v$_2$,0) & (0,v$_3$,0)$\phi$ & (0,v$_3$,0) & (0,v$_3$,0)$\phi$ \\
$\Gamma_3$  & (u$_1$,0,0) & (u$_1$,0,0)$\phi$ & (u$_2$,0,w$_2$) & (u$_2$,0,-w$_2$)$\phi$ & (u$_3$,0,w$_3$) & (u$_3$,0,-w$_3$)$\phi$ \\
$\Gamma_4$ & (0,v$_1$,w$_1$) & (0,-v$_1$,w$_1$)$\phi$ & (u$_2$,0,w$_2$) & (-u$_2$,0,w$_2$)$\phi$ & (u$_3$,0,w$_3$) & (-u$_3$,0,w$_3$)$\phi$ \\
\end{tabular}
\caption{Basis vectors for symmetry-related Co ions of the cross-tie and spine sites belonging to irreducible representations of the little group $G_k$, Cmca with $\textbf{k}=(0,\delta,0)$, $\phi=exp(i \pi \textbf{k})$. Labeling of the atoms follows the conventions shown in Fig.~\ref{fig:structure}. Labeling of the irreducible representations follow Kovalev's notation~\cite{Kovalev}.}
\label{basis_vector_icm}
\end{table*}

Symmetry analysis of the magnetic structures for the HT$_{\rm inc}$, HT$_{\delta =\frac{1}{2}}$ and the LT$_{\rm F}$ phases have been performed using representation analysis. The labeling of atomic positions for cross-tie and spine sites is shown in Fig.~\ref{fig:structure}. Analysis of the magnetic configurations for atomic sites generated from C-centering translation of the labeled positions is not explicitly reported but can be easily calculated by including the phase factor $2 \pi \textbf{k} \cdot \textbf{T} $ where $\textbf {T} = (\frac{1}{2}, \frac{1}{2}, 0)$. For the HT structures (\textbf{k}=\textbf{k$_8$} in Kovalev's notation), only the screw-axis along \textit{y}, mirror plane perpendicular to \textit{x}, and b glide plane perpendicular to the \textit{z}-axis are valid operators in the little group. The loss of symmetry operators indicate that the four equivalent positions for the Co spine sites split into two orbits, Co(1-1) with Co(1-2) and Co(2-1) with Co(2-2). The two Co cross-tie sites, Co(c-1) and Co(c-2) belong to the same orbit. Basis vectors of the HT magnetic structures are reported in Table~\ref{basis_vector_icm} for the four irreducible representations labeled $\Gamma_i (i=1,4)$, all mono-dimensional. In agreement with the results of Chen {\it et al.},\cite{Chen} we found that only $\Gamma_3$ fits the experimental data with modes along the \textit{a}-axis. There is no canting along \textit{c} within the resolution of our experiment.

Although the Co spine sites split into two orbits, co-representation analysis including the complex conjugation operator constrain the moment on both orbits to be the same. However, there are no symmetry constraints as such for the phase differences. Therefore, in addition to a global phase factor, set to be 0 on the first Co(c-1) site, there are two phases to be refined describing the relative phasing of the two spin-density waves on the Co(1-1)/Co(1-2) and Co(2-1)/Co(2-2) orbits with respect to the Co(c) wave. These phases, in units of $2\pi$ radians, were found to be 0.18(5) and 0.09(5) at 8.2~K where the propagation vector locks at the commensurate value $\delta=\frac{1}{2}$. The results of Ref.~\onlinecite{Chen} can be reproduced by adding a global phase of $\pi/2$ to the values reported here.

On warming the sample the phase values increase and at 10.7~K, where $\delta = 0.54$, they reach 0.27(5) and 0.15(5), for the two spin-density waves respectively. The magnetic structure at 9.7~K in the HT$_{\rm inc}$ phase is presented on Fig.~\ref{fig:structure}. It is similar to that found in the HT$_{\delta=\frac{1}{2}}$ phase, the low-magnetic moments on cross-tie sites are located in layers for which adjacent spine chains are aligned antiferromagnetically. At 8.2~K, the amplitude of the waves were found to be 1.1(2) and 2.99(2)~$\mu_{\rm B}$ on the Co cross-tie and Co spine sites, respectively. Whereas the expected value for high spin Co$^{2+}$ in octahedral configuration (spin only) is achieved on the spine site, there is a marked reduction of the moment on the cross-tie site indicating large fluctuations even in the commensurate HT$_{\delta=\frac{1}{2}}$ phase. At 10.7~K, the amplitudes of the waves decreases to 0.90(18) and 1.99(4)~$\mu_{\rm B}$, respectively.

\begin{table*}[tb]
\begin{tabular}{c|cccccc}
 & & & Atoms & & & \\
Irreps. & Co(c-1) & Co(c-2) & Co(1-1) & Co(1-2) & Co(2-1) & Co(2-2) \\
\hline $\Gamma_1$ & (u$_1$,0,0) & (-u$_1$,0,0) & (0,v$_2$,0) & (0,-v$_2$,0) & (0,v$_2$,0) & (0,-v$_2$,0) \\
$\Gamma_2$ & - & - & (0,v$_2$,0) & (0,v$_2$,0) & (0,-v$_2$,0) & (0,-v$_2$,0) \\
$\Gamma_3$ & (u$_1$,0,0) & (u$_1$,0,0) & (u$_2$,0,w$_2$) & (u$_2$,0,-w$_2$) & (u$_2$,0,w$_2$) & (u$_2$,0,-w$_2$) \\
$\Gamma_3$ & (u$_1$,0,0) & (u$_1$,0,0) & (u$_2$,0,w$_2$) & (u$_2$,0,-w$_2$) & (u$_2$,0,w$_2$) & (u$_2$,0,-w$_2$) \\
$\Gamma_4$ & - & - & (u$_2$,0,w$_2$) & (-u$_2$,0,w$_2$) & (-u$_2$,0,-w$_2$) & (u$_2$,0,-w$_2$) \\
$\Gamma_5$ & (0,v$_1$,w$_1$) & (0,v$_1$,-w$_1$) & (0,v$_2$,0) & (0,v$_2$,0) & (0,v$_2$,0) & (0,v$_2$,0) \\
$\Gamma_6$ & - & - & (0,v$_2$,0) & (0,-v$_2$,0) & (0,-v$_2$,0) & (0,v$_2$,0) \\
$\Gamma_7$ & (0,v$_1$,w$_1$) & (0,-v$_1$,w$_1$) & (u$_2$,0,w$_2$) & (-u$_2$,0,w$_2$) & (u$_2$,0,w$_2$) & (-u$_2$,0,w$_2$) \\
$\Gamma_8$ & - & - & (u$_2$,0,w$_2$) & (u$_2$,0,-w$_2$) & (-u$_2$,0,-w$_2$) & (-u$_2$,0,w$_2$) \\
\end{tabular}
\caption{Basis vectors for symmetry-related Co ions of the cross-tie and spine sites belonging to irreducible representations of the space group Cmca.
Labels of the atoms follow the convention shown in Fig.~\ref{fig:structure}.
Labeling of the irreducible representations follows Kovalev's notation.
The absence of modes for irreducible representations not appearing in the decomposition of the magnetic representation is indicated by the ``-" sign.}
\label{basis_vector_cm}
\end{table*}

Next we consider the LT$_{\rm F}$ magnetic phase. Symmetry analysis is now performed in the group Cmca since all symmetry operators belong to the little group for $\textbf{k}=0$. There are eight one-dimensional representations in this case and the atomic positions for each Co site belong to the same orbit. The projected basis vectors are shown in Table~\ref{basis_vector_cm}. Although the transition to the ferromagnetic state is first order and Landau theory does not restrain the magnetic modes to belong to the same irreducible representation in this case, the data below 5.7~K are consistent with a ferromagnetic structure described by a single representation, $\Gamma_3$. Similarly to the HT structure, the magnetic moments are directed along the \textit{a}-axis. A weak canting along \textit{c}, allowed by symmetry, is not excluded since the data shows a weak hump around the (110) Bragg position. However, the contribution is too small to extract a sizeable component along this direction. The extracted moments at 2~K are 1.81(4)~$\mu_{\rm B}$ and 3.04(2)~$\mu_{\rm B}$ on the cross-tie and spine sites respectively. Between 2~K and 5.7~K, there are no noticeable variations of the magnetic moment. Again, the moment on the cross-tie site is largely reduced with respect to the expected spin contribution of 3~$\mu_{\rm B}$.

\subsection{Magnetic structure under magnetic field}
\begin{figure}[tb]
\includegraphics[width=\columnwidth]{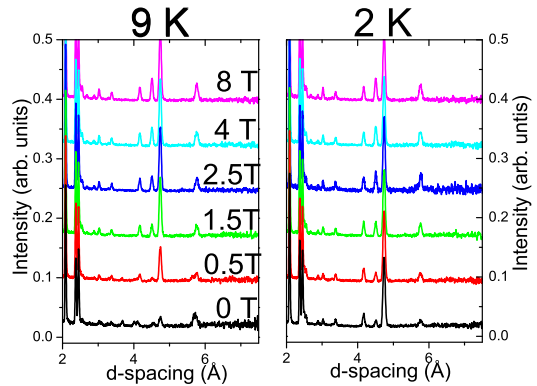}
\caption{\label{fig:fielddata} (color online) Neutron powder diffraction data at 9~K (left) and 2~K (right) under magnetic fields of 0, 0.5, 1.5, 2.5, 4.0 and 8.0~T. Data from a bank of detectors situated at the scattering angle $35^{\circ}$ are shown. Individual curves are offset arbitrarily for display purposes.}
\end{figure}

To gain further insight into the magnetic behavior of \CVO, powder neutron diffraction experiments in magnetic fields have been conducted. First, the field was applied at 9~K where the system is in the HT$_{\rm inc}$ phase. The data are presented in Fig.~\ref{fig:fielddata}. At 0.5~T, we observe a shift in the positions of magnetic Bragg peaks, indicating a change in propagation vector of 0.02 even at this moderate field value. Simultaneously, new magnetic Bragg peaks appear, indexed at $\textbf{k}=0$. In particular, we notice a strong increase of the (021) reflection. With further increase of the magnetic field, the incommensurate magnetic state completely disappears, leaving an essentially pure ferromagnetic pattern, similar to that observed for the LT$_{\rm F}$ structure in zero-field. However, the relative intensities of several magnetic Bragg peaks are different from the zero-field data and vary with increasing magnetic field: the (111), (020), (021) and (220) peaks increase continuously between 1.5 and 8~T, suggesting a possible spin-reorientation.

\begin{figure}[tb]
\includegraphics[width=\columnwidth]{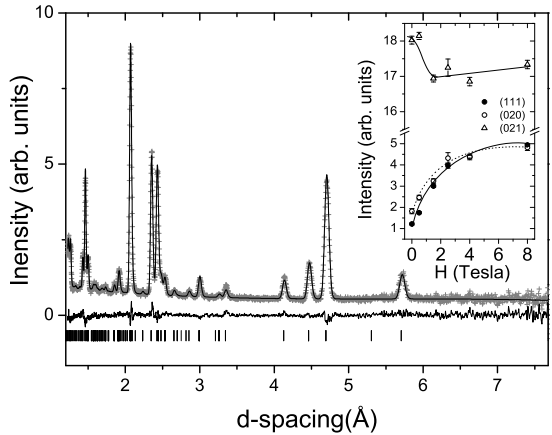}
\caption{\label{fig:rietveld} Rietveld refinement of the nuclear and magnetic structure of \CVO\ at 2~K, 8~T. Experimental data points and results of the refinement are shown as grey crosses and a thin solid line respectively. The lower solid line represents the difference curve between experimental and calculated pattern. The positions of Bragg reflections are indicated by tick marks. Inset: Integrated intensities of selected reflections as a function of magnetic field. Curves are guides to the eye.}
\end{figure}

Applied field neutron diffraction patterns collected in the LT$_{\rm F}$ state at 2~K are presented in Fig.~\ref{fig:fielddata}. No additional magnetic reflections appear, indicating that the $\textbf{k}=0$ structure is stable under application of magnetic fields up to 8~T. However, we observe a redistribution of the scattering intensities (see inset of Fig.~\ref{fig:rietveld}) : the (111) and (020) peaks strongly increase with field. In addition, the integrated intensity of the (111) reflection, increases from being smaller than the (020) integrated intensity at 0~T, to slightly higher at 8.0~T. At the same time, the intensity of the strongest magnetic peak (020) does not vary up to 0.5~T, and only decrease slightly for higher field values.

We initially tried to model the diffraction data in magnetic fields with an identical arrangement to the zero-field LT$_{\rm F}$ structure, allowing only the magnetic moment on both sites to vary. This model led to an enhancement of the magnetic moments on the Co cross-tie site but failed overall to accurately reproduce the experimental data, resulting in a poor agreement factor for the magnetic phase of ${\rm R}({\rm F}^2)_{Mag}=16.80$\%. Subsequent analysis showed that the data are best reproduced with a model in which the spins of both Co sites are allowed to rotate in the \textit{ac}-plane. This corresponds to a structure where modes belonging to the irreducible representation $\Gamma_3$ along \textit{x} and $\Gamma_7$ along \textit{z} are mixed. The agreement with the experiment is very good, as shown in Fig.~\ref{fig:rietveld} and a reliability factor of ${\rm R}({\rm F}^2)_{Mag}=8.76$\% is obtained for Rietveld refinement of the data collected at 8~T. 

\begin{figure}[tb]
\includegraphics[width=\columnwidth]{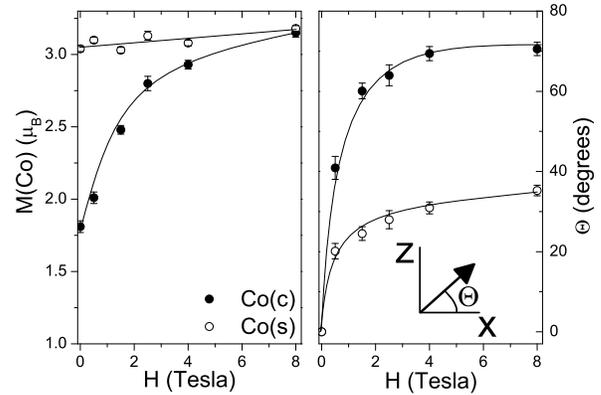}
\caption{\label{fig:hparameters} Dependence of the low-temperature magnetic structure parameters on magnetic field, extracted from Rietveld refinements. Left panel: magnetic moments on the cross-tie site (solid circle) and spine site (open circle). Right panel: angular deviation of spins in the \textit{ac}-plane for the cross-tie site (solid circle) and spine site (open circle). $\Theta$=0 corresponds to a collinear structure with spins lying along the \textit{a}-axis. Curves are guides to the eye.}
\end{figure}

Two essential characteristics can be extracted from the refinement, these are summarized in Fig.~\ref{fig:hparameters}.
Firstly, the magnetic moment on the Co cross-tie site increases rapidly with application of the magnetic field and saturates at 3.15(3) $\mu_{\rm B}$ at 8~T. 
In zero field the moment magnitude is only 57\% of the fully saturated value observed in an applied field of 8~T.
The magnetic moment on the Co spine site barely changes with field and is refined at 3.18(2)~$\mu_{\rm B}$ at~8 T.
Both values agree well with the expected spin-only contribution for high-spin octahedral Co$^{2+}$.
It is clear that one must be cautious when reporting analysis of this kind due to the intrinsic limitations of the experiment with polycrystalline samples. 
However, semi-quantitative information can be accessed because a large Q-range is probed, a condition not realized for single-crystal instruments due to the geometrical constraints imposed by a cryomagnet. 

Secondly, the refinement clearly indicates a reorientation of the spins in the \textit{ac}-plane in an applied field.
With powder experiments, extracted moment directions are averaged over all possible orientations of the field with respect to the crystal axes, including the \textit{b}-axis.
From magnetization measurements,~\cite{Szymczak, Balakrishnan, correction, WilsonHFM} it has been shown that the \textit{b}-axis of this material is clearly the hard axis. 
The magnetization is an order of magnitude smaller than that from the \textit{c}-axis, which is a further order of magnitude smaller than that of the \textit{a}-axis.
For an ideal ferromagnetic material for which \textit{a} and \textit{c} are equally easy directions and the \textit{b} axis is perfectly hard, one can expect that the magnetic moments will be fully polarized along the applied field irrespective of the field direction in the \textit{ac}-plane.
For such an idealised material, a powder measurement integrated over all possible orientations would show an average moment direction pointing at $45^\circ$~degrees in the \textit{ac}-plane.
In \CVO, however, the moment orientation on the Co spine site is slightly less than the expected 45$^{\circ}$ value, while the moment orientation angle of the cross-tie site is much higher than expected.
From the zero field data presented here, and the magnetisation data presented elsewhere,~\cite{WilsonHFM} it is clear that the \textit{a}-axis is the easy axis of the system. 
However, on application of a magnetic field, the \textit{c}-axis becomes an easier direction than the \textit{a}-axis for the Co cross-tie site.

In the LT$_{\rm F}$ phase, the cross-tie moments are not as large as expected, and lie along the \textit{a}-axis parallel to the fully saturated spine site moments. 
In an applied field of 8~T, the cross-tie moments have reached the expected saturation value and point more along the \textit{c}-axis than the \textit{a}-axis. 
The spine site moments have also reorintated to have a larger component along the \textit{c}-axis than in zero field, but not as much as the cross-tie moments. 
This suggests that the single ion anisotropies for the two sites are different and that the application of a field may influence the two moments in different ways.
 
 
One should discuss this behavior qualitatively and in the limits of the model approximations, in particular the hypothesis of a perfectly hard direction along \textit{b} (obviously not fulfilled). 
Overall, these results are in good agreement with magnetization measurements showing that \textit{a} and \textit{c} are definitely easy-directions in \CVO.\cite{Balakrishnan, correction, WilsonHFM} 
The fact that the moment on the cross-tie site increases rapidly from the reduced 1.81(4)~$\mu_{\rm B}$ value is also in agreement with the fully polarized ferromagnetic structure observed in magnetization measurements in moderate magnetic fields.  
The slightly different anisotropies for both Co sites, suggested by our neutron experiments, could explain the different magnetic behaviors observed when the field is applied along \textit{c} or \textit{a}.

\section{Conclusion}
Powder neutron diffraction studies have been performed in zero and applied magnetic fields of up to 8~T on the \Kag\ staircase oxide \CVO.
The magnetic structures in the system have been refined using the FullProF program.
The system has two main phases below 11~K: a spin density wave structure with $\textbf{k}=(0,\delta,0)$, where $\delta$ varies around $\frac{1}{2}$ above 6~K and a commensurate $\textbf{k}=0$ ferromagnetic structure below.
In an applied field, the moment magnitude of the cross-tie site (one of two Co sites) is found to increase from a much reduced value to the expected saturation value. 
During this increase, a reorintation of all of the Co moments is observed.
These results indicate that the application of a magnetic field relieves the frustration on the cross-tie site and that the two Co sites have different anisotropies.

\begin{acknowledgments}
We are grateful to B F{\aa}k for useful discussions and to the EPSRC for financial support.
\end{acknowledgments}

\bibliography{copaper}

\begin{thebibliography}{13}
\bibitem[{cor()}]{correction}
\bibinfo{note}{Reference 9 uses a different notation to this paper and to
  reference 2 for labelling axes, the space group Acam was used so the c and a
  axes are swapped.}
\end{thebibliography}

\end{document}